\documentclass[twocolumn,showpacs,amsmath,amssymb]{revtex4}
\usepackage{graphicx}

\begin{document}

\title{Revealing Majorana Fermion states in a superfluid of cold atoms subject to a harmonic potential}

\author
{Tao Zhou$^{1,2}$ and Z. D. Wang$^{1}$}

\affiliation{$^{1}$Department of Physics and Center of Theoretical and Computational Physics, The University of Hong Kong, Pokfulam Road, Hong Kong, China\\$^{2}$College of Science, Nanjing University of Aeronautics and Astronautics, Nanjing 210016, China}

\date{\today}
\begin{abstract}
We here explore Majorana Fermion states in an $s$-wave superfluid of cold atoms in the presence of
spin-orbital coupling and an additional harmonic potential. The superfluid boundary is induced by a harmonic trap.
Two locally separated Majorana Fermion states are revealed numerically based on the self-consistent Bogoliubov-de Gennes equations.
The local density of states are calculated, through which the signatures of Majorana excitations may be indicated experimentally.
\end{abstract}
\pacs{71.10.Pm, 03.67.Lx, 74.90.+n}
 \maketitle

The Majorana Fermion (MF), a particle that
is its own antiparticle, was originally proposed by Majorana many years ago~\cite{maj}. For the past decade, the MFs have attracted tremendous attentions due to their exotic properties. Especially manipulating MFs
is one promising way to realize the non-Abelian statistics that may have potential applications in topological quantum computation~\cite{naya}.  For condensed matter systems, the MFs were first predicted to emerge in the Pfaffian fractional quantum Hall state~\cite{moor}. Later it was proposed that the MFs exist in the $p+ip$ superconducting system associated with the zero energy bound states in the vortex core~\cite{read,iva,sum}.
While searching for the spin-triplet $p$-wave superconducting materials is of a great challenge.
Besides the real material, the $p+ip$ superfluid (SF) pairing state was proposed to be artificially created in the cold atom systems with the $p$-wave Feshbach resonance~\cite{tew}. However, p-wave molecules are not stable so that the proposal is also difficult to implement~\cite{inada}. Recently, more attention has been paid to topological systems with the spin-orbital interaction~\cite{lfu,sato,jliu,jsau,lin,sau,zhu,huhui}.
This kind of system with an $s$-wave pairing may be equivalent to the $p+ip$ superconductor~\cite{lfu,sato,jliu}.
Experimentally, tremendous efforts have been made to search for MFs. Very recently, several groups have reported the signatures of MFs in various systems~\cite{wil,rok,den,das,mou}, but a definite demonstration for the existence of MFs is still awaited.

It has been indicated that both the spin-orbital coupling and the $s$-wave pairing SF state can be artificially created in cold atom systems. The spin orbital coupling can be generated through spatially varying laser fields~\cite{zhu,juz,szhu}. The $s$-wave pairing state is much more stable than the $p$-wave one and has been realized~\cite{bou}. Besides the above conditions, the realization of MFs require the presence of the Zeeman field and vortex.  A feasible scheme to realize a controllable Zeeman field in cold atoms was proposed in Ref.~\cite{zhu}.
In the present work, instead of the realization of MFs in the vortex core, we elaborate for the first time that the MF states may also be induced in a confined harmonic potential, noting that the position and motion of the confined potential may be controlled more easily in the cold atoms.
Based on the Bogoliubov-de Gennes (BdG) equations, the SF order parameters are calculated self-consistently. Our numerical results illustrate the existence of the MF states. The distinct features of the MF states are discussed in detail. 

We start from a Hamiltonian including the spin-orbital coupling and the SF pairing term, which reads
\begin{equation}
H=H_t+H_\Delta,
\end{equation}
Here $H_t$ includes the tight-banding part of the model with the spin-orbital interaction and the spin polarization term, given by~\cite{qi}
\begin{eqnarray}
H_t=&\sum_{\bf i}[\psi^{\dagger}_{\bf i}\dfrac{{\sigma_3}-i{\sigma_1}}{2}\psi_{{\bf i}+\hat{x}}+\psi^{\dagger}_{\bf i}\dfrac{{\sigma_3}-i{\sigma_2}}{2}\psi_{{\bf i}+\hat{y}}+h.c.]\nonumber\\
&+\sum_{\bf i} \psi^{\dagger}_{\bf i}(U_{\bf i}\sigma_0+h\sigma_3)\psi_{\bf i}
\end{eqnarray}
with $\psi_{\bf i}=(\psi_{{\bf i}\uparrow},\psi_{{\bf i}\downarrow})^\mathrm{T}$, where $\sigma_n$ are the identity (n=0) and Pauli matrix $(n=1,2,3)$, respectively, $h$ represents an effective Zeeman field.
While $H_\Delta$ is the SF pairing term, expressed as
\begin{equation}
H_\Delta=\sum_{\bf i}(\Delta_{\bf i} \psi^\dagger_{\bf i} i\sigma_2 \psi^\dagger_{\bf i}+h.c.).
\end{equation}

The above Hamiltonian can be diagonalized by solving the
BdG equations self-consistently,
\begin{equation}
\left(
\begin{array}{cc}
 H_t({\bf r}) & \Delta({{\bf r}})\sigma_3  \\
 \Delta^{*}({{\bf r}})\sigma_3 & -\sigma_2 H^{*}_{t}({\bf r}) {\sigma_2}
\end{array}
\right)  \begin{array}{c}
\Psi^{n}({\bf r})
\end{array}
 =E_n \begin{array}{c}
\Psi^{n}({\bf r})
\end{array}.
\end{equation}
For the $N\times N$ lattice size, the dimensionality
 of the Hamiltonian matrix is $4N\times 4N$.
$\Psi^{n}({\bf r})$ is a $4N$ order column vector with $\Psi^{n}({\bf r})=(u^{n}_{{\bf r}\uparrow},u^{n}_{{\bf r}\downarrow},v^{n}_{{\bf r}\downarrow},v^{n}_{{\bf r}\uparrow})^{\mathrm{T}}$.

The order parameters $\Delta({\bf r})$ are determined self-consistently,
\begin{eqnarray}
\Delta({\bf r})=\frac{V}{4}\sum_n
(u^{n}_{{\bf r}\uparrow}v^{n*}_{{\bf r}\downarrow}+u^{n}_{{\bf r}\uparrow}v^{n*}_{{\bf r}\downarrow})\tanh
(\frac{E_n}{2K_B T}),
\end{eqnarray}
with $V$ the pairing strength.

The on-site particle number with spin $\sigma$ is calculated after diagonalizing the BdG Hamiltonian,
\begin{equation}
\langle n_{{\bf i}\sigma}\rangle=\sum_n \mid u^{n}_{{\bf i}\sigma} \mid^2 f(E_n),
\end{equation}
where $f(x)$ is Fermi distribution function. Then we can obtain the on-site particle number $n_{\bf i}=\langle n_{{\bf i}\uparrow}\rangle+\langle n_{{\bf i}\downarrow}\rangle$ and the site-dependent magnetization $m_{\bf i}=1/2(\langle n_{{\bf i}\uparrow}\rangle-\langle n_{{\bf i}\downarrow}\rangle)$, respectively.
The local density of states (LDOS) can be calculated as:
\begin{equation}
\rho_{\bf i}(\omega)=\sum_n [\mid u^{n}_{{\bf i}\uparrow} \mid^2 \delta(E_n-\omega)+\mid v^{n}_{{\bf i}\downarrow} \mid^2 \delta(E_n+\omega)].
\end{equation}
The delta function $\delta(x)$ is taken as $\delta=\Gamma/[\pi(x^2+\Gamma^2)$], with the
quasiparticle damping $\Gamma=0.01$.

 We first analyze theoretically the duality between the present model with the $p$-wave superconductors.
  For the uniform case with $U_i=0$, the Hamiltonian can be expressed in the momentum space,
 \begin{equation}
 H=H_{t{\bf k}}+\sum_{\bf k}(\Delta_{\bf k} e^{i\phi_{\bf k}}\psi_{{\bf k},\uparrow}\psi_{{-\bf k},\downarrow}+h.c.),
 \end{equation}
where $H_{t{\bf k}}$ is the tight-banding term given by
\begin{equation}
H_{t\bf k}=\sum_{\bf k}h({\bf k})(\psi^{\dagger}_{{\bf k}\uparrow}\psi_{{\bf k}\uparrow}-\psi^{\dagger}_{{\bf k}\downarrow}\psi_{{\bf k}\downarrow})+\sum_{\bf k}[g({\bf k})\psi^{\dagger}_{{\bf k}\uparrow}\psi_{{\bf k}\downarrow}+h.c.],
 \end{equation}
 with $h({\bf k})=(m+\cos k_x+\cos k_y)$ and $g({\bf k})=\sin k_x+i \sin k_y$.
 Defining two spinless quasiparticle operators $c_{\bf k}=\frac{1}{\sqrt{2}}(\psi^\dagger_{{\bf k}\uparrow}+ e^{i\phi_{\bf k}}\psi_{{\bf -k}\downarrow})$ and $d_{\bf k}=\frac{i}{\sqrt{2}}(\psi^\dagger_{{\bf k}\uparrow}-e^{i\phi_{\bf k}}\psi_{{\bf -k}\downarrow})$,
the Hamiltonian can be rewritten in the spinless representation:
\begin{eqnarray}
H=\sum_{\bf k}\Delta_{\bf k}C^{\dagger}({\bf k})\sigma_0 C({\bf k})+\sum_{\bf k}h({\bf k})C^{\dagger}({\bf k})\sigma_3 C({\bf k})
\nonumber\\+\sum_{\bf k}[g({\bf k})e^{i\phi_{\bf k}}C^{\dagger}({\bf k})C^{\dagger}({-\bf k})+h.c.],
\end{eqnarray}
with $C({\bf k})=[(d({\bf k}),c({\bf k})]^{\mathrm{T}}$.
One can find that the system is dual to the $p+ip$ system with the pairing term $\widetilde{\Delta}=\sin k_x+i\sin k_y$.
Moreover, the spin-orbital interaction is transformed to the pairing term.
  This dual transformation makes it possible to realize the MFs in the spin-orbital system.

The $H_t$ in Eq.(2) with $U_{\bf i}=0$ is an effective model to describe the topological insulator~\cite{qi}.
The topological properties are determined by $h$
 with the non-zero Chern number for $0<\mid h\mid<2$. Fig.1(a) illustrates the chiral edge states of the model by considering the open boundary condition along $x$ direction ($L_x=50$) and periodic boundary condition along $y$-direction with $h=-0.2$. As is seen, the energy gap closes at $k_y=\pi$ with the zero energy mode being localized at the edge $(i_x=0,50)$,
  suggesting the existence of the edge state.
    On the other hand, if periodic boundary condition is taken for both directions, the Hamiltonian can be expressed as $2\times2$ matrix in the momentum space [Eq.(9)].
 The two energy bands, obtained by diagonalizing the above $2\times2$ matrix, are plotted in Fig.1(b), representing  the energy spectrum in the bulk.
There are two bands with the gap $0.4$ and the energy bandwidth $2$ ($\varepsilon\ni[-2.2,-0.2]\cup[0.2,2.2]$).
If the on-site potential $U_{\bf i}$ with $0.2<U_{\bf i}<2.2$ is chosen, the Fermi energy would cross the lower energy band and the system becomes metal-like. Then, the SF pairing might occur when an additional pairing potential term is taken into account.

\begin{figure}
\centering
  \includegraphics[width=8.2cm]{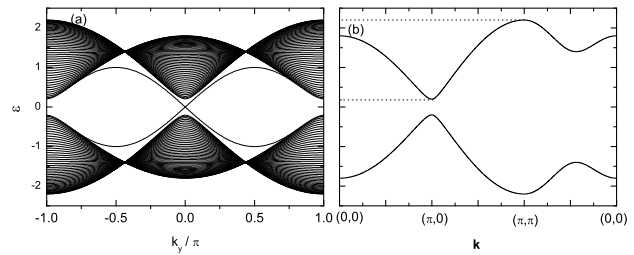}
\caption{(a) The energy spectrum of the Hamiltonian Eq. (2) with an open boundary condition along the $x$-direction. (b) The energy band in the momentum space with periodic condition. }
\end{figure}

The presence of the vortex is likely a key ingredient but a great challenge for probing the MFs and non-Abelian statistics.
Usually, the vortex is induced by the magnetic field, characterized as a $2\pi$ winding of the phase and the vanishing of the order parameter in its center.
In the present work, we will not consider the field-induced vortex, rather take into account
an harmonic potential in the form of  $U_{\bf i}=U_0 \mid({\bf r_i}-{\bf r_0})\mid^2$.
We can conclude from the band structure shown in Fig.1(b) that
the SF order parameter is nearly zero in the trap center and increases when away from the center.
It will reach the maximum value when the potential $U_i$ crosses the energy band. Two SF boundaries are expected to exist when the on-site potential crosses the band edges.
The MF excitation may appear at the SF boundaries.

Now let us illustrate numerically the existence of the MF states based on the above proposal. In the following, we use $h = -0.2$, $U_0= 0.01$ and the
pairing potential $V = 2$.
The numerical calculation is performed on a $36\times 36$ square lattice with the periodic boundary condition. The trapping center locates at ${\bf r_0}=(18,18)$.

The self-consistent results of the BdG equations are presented in Fig.2. Fig.2(a) displays
the spatial distribution of the gap magnitudes and their phases (denoted by arrows).
The spatial distribution of the particle number and the site-dependent magnetization are presented in Figs.2(b) and 2(c), respectively.
The two-dimensional cut of the gap magnitude, the particle number, and the site dependent magnetism are plotted in Fig.2(d).
As is seen, the gap is nearly zero at the trap center and increases when it is away from the center. It reaches the maxima at about $\mid {\bf r_i}-{\bf r_0} \mid=11$. Then the gap decreases and becomes nearly zero at the border. The SF region forms a ring with the gap magnitudes being nearly isotropic around the ring. The phases of the SF order parameters
 change from 0 to $2\pi$ around the ring. Thus the self-consistent results for the order parameters are somewhat similar to the case of the vortex state for superconducting materials in the magnetic field. Actually, here the presence of the phase winding is due to the presence of the Zeeman field, which is also an effective vertical magnetic field.

\begin{figure}
\centering
  \includegraphics[width=8cm]{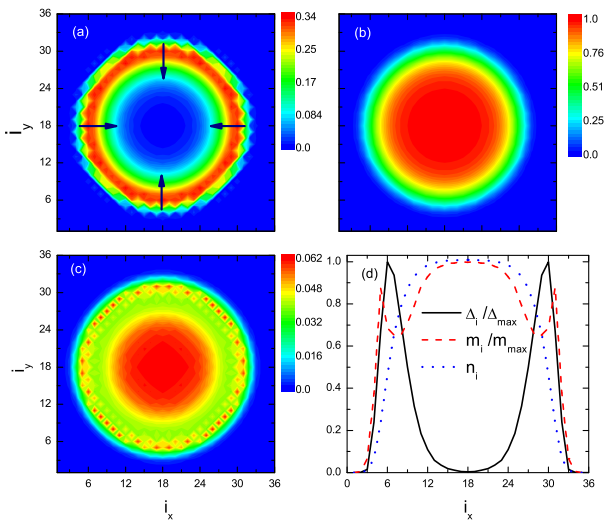}
\caption{(Color online) (a) The numerical results for the magnitudes and phases (shown by arrows) of the pairing order parameters.
(b)and (c) are the particle number and magnetism, respectively. (e) The two dimensional cut of the pairing magnitude, particle number and magnetism along $i_y=18$, respectively.}
\end{figure}

  The competition between the SF and ferromagnetism is elucidated in Figs.2(b-d), which is a key point for the zero mode discussed below. When
  $\mid {\bf r_i}-{\bf r_0} \mid<4.6$ ($U_{\bf i}<0.2$), the on-site energy is smaller than the minimum excitation energy, and thus the system behaves as a ferromagnetic insulator with the particle number per site being fixed to near 1.0.
  The magnetization reaches a maximum and the pairing order parameter is negligibly small.
  As $4.6<\mid {\bf r_i}-{\bf r_0} \mid<15$ ($0.2<U_{\bf i}<2.2$), the Fermi energy crosses the energy band. Therefore, the lower band becomes unfilled, so that the average on-site particle number and the ferromagnetic order decrease. As a result, the SF order shows up. We can also see that the evolution of ferromagnetism is non-monotonous, namely, it reaches the minimum as the pairing order is of maximum and recover to a local maxima value as the pairing order decreases, indicating the competition between SF order and ferromagnetism.
As $\mid {\bf r_i}-{\bf r_0}\mid>15$, the on-site energy is larger than the maximum-excited energy of the band. Both the particle number and the order parameters decrease to zero.

As presented in Fig.2, when the on-site energy is small or very large, the particle number is fixed to be 2 or 0, and thus an insulating gap should exist. The insulating gap disappears when the on-site energy crosses the energy band. The system becomes metal-like and
the SF pairing gap shows up.
We can expect two effective insulator-SF boundaries inside and outside the SF ring. The zero mode is expected to exist at the boundaries. We diagonalize the Hamiltonian
and all of the 5184 eigen-values are plotted in Fig.3. As is seen, the eigen-values are nearly continuous
but a kink occurs at the zero energy. Enlarging the low energy part in the inset of Fig.3, the eigen-values are discontinuous. Intriguingly, our numerical calculation reveals the existence of a zero energy fermionic mode. This zero energy mode is protected by an energy gap about 0.027, so that
it is robust and can hardly be excited  by local perturbations.

\begin{figure}
\centering
  \includegraphics[width=7.5cm]{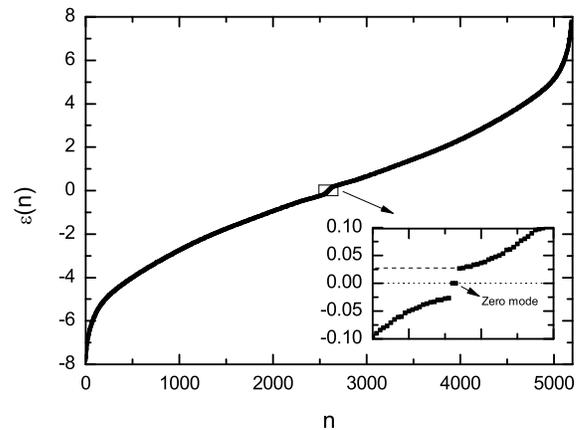}
\caption{The eigen-values for the Hamiltonian matrix. Inset: The replot of the rectangle part in the main figure.
}
\end{figure}

As is known, the creation/annihilation operator of the captioned zero mode is of self-hermitian~\cite{read}, namely, there may be two MFs associated with this zero mode, with the corresponding operators
being given by
$\gamma_{1}=(C+C^{\dagger})/\sqrt{2}$ and $\gamma_{2}=i(C^{\dagger}-C)/\sqrt{2}$,
where $C^{\dagger}=\sum_{i\sigma}(u_{i\sigma}\psi^{\dagger}_{i\sigma}+v_{i\sigma}\psi_{i\sigma})$, which can be
obtained from the zero energy eigen-vector.
In this sense, we are able to study numerically the MF states
by solving the BdG equations. The
spatial distribution of the two MF states $\rho_{1,2}$ with the units of $\rho_{max}$ are presented in Figs.4(a) and 4(b), respectively. Since two MFs at the same location shall annihilate to an ordinary fermion, only locally separated MFs are meaningful.
It should be insightful to eliminate the overlap part of the two MF states by looking into the difference of the two MFs ($\mid\rho_1-\rho_2\mid$), as presented in Fig.4(c). The two-dimensional cut of $\rho_{1}-\rho_2$ along $i_y=18$ is plotted in Fig.4(d).

\begin{figure}
\centering
  \includegraphics[width=8cm]{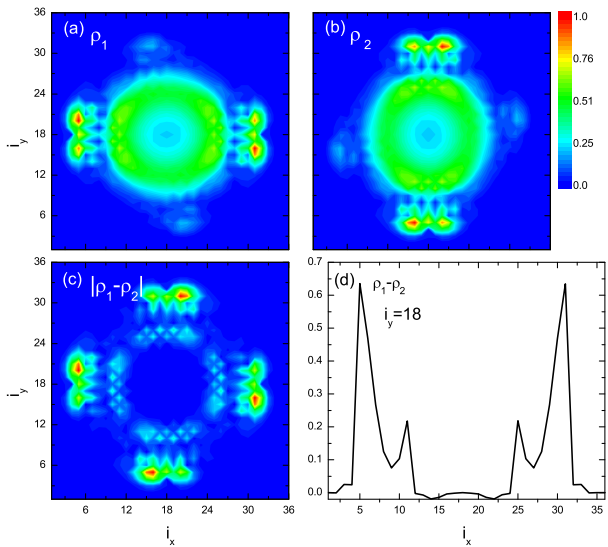}
\caption{(Color online) The intensity plot of the spatial distributions of (a) $\gamma_1$, (b) $\gamma_2$. (c) The difference of the probability distributions between $\gamma_1$ and $\gamma_2$. (d) The two-dimensional cut of $\rho_1-\rho_2$ along $i_y=18$.
}
\end{figure}

The two separated MF states may be identified from the results presented in Fig.4.
They appear at the SF-insulator boundaries with maximum probabilities being at the horizontal (vertical) directions for $\gamma_1$ $(\gamma_2)$.
One significant feature revealed by Fig.4 is the non-local behavior for both states.
The MF states are symmetric about the trapping center, and thus both of them should have two equivalent maxima.
Inside the SF ring, there exists a certain extent of overlap between these two states; while around outsider boundary, there is almost no overlap between the two states.
This overlap behavior is a distinct feature and would not occur for the usual MFs in the well-separated vortex cores. As is known, for the usual vortex bounded MF, the wave function decreases exponentially away from the vortex core. For the present case, the overlap of the MF states seems more interesting and may be controlled by the trap potential.


\begin{figure}
\centering
  \includegraphics[width=8.3cm]{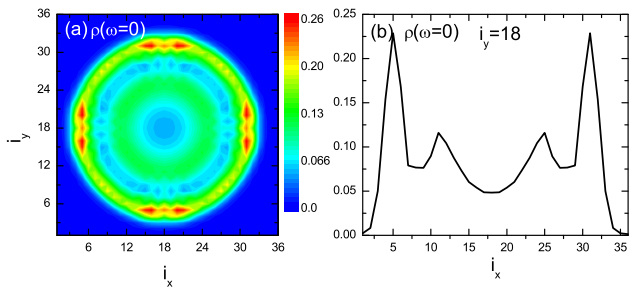}
\caption{(Color online) (a) The intensity plot of the zero energy LDOS. (b) The two dimensional cut of the zero energy LDOS.
}
\end{figure}

It is also insightful to look into the LDOS for investigating the existence and distribution of the above zero mode. The intensity plot of the zero energy LDOS spectra is plotted in Fig.5(a). Its two-dimensional cut for $i_y=18$ is plotted in Fig.5(b). Generally the zero energy LDOS could describe the excitations of low energy quasiparticle fermions. Due to the existence of the quasiparticle damping, it is contributed not only the exact zero energy quasiparticles, but also some low energy fermionic states. For the present case, since there exists an energy gap between the zero energy state and the lowest excited state, as is shown in Fig.3, the contributions from non-zero energy state should be rather small.
As a result, here the LDOS spectra may be qualitatively consistent with the distributions of the two MF states. This point could be seen by comparing Fig.4 and Fig.5. Especially, from the two dimensional cut of the spectra shown in Fig.4(d) and Fig.5(b), our results indicate that the zero energy LDOS are qualitatively the same as the results of $\rho_1-\rho_2$. Therefore, the indication of MF modes may be provided through observing experimentally the LDOS spectra. This may establish a useful link for theoretical analysis and experimental observations.

In summary, we have proposed and elaborated a new scenario to realize the MF states in the fermionic cold atom systems. 
The numerical results have shown that
the SF region forms a ring with the MF states being mainly around the boundaries of the ring.
It is rather promising that the presence of MF zero modes
may be detected experimentally through the LDOS spectra.

This work was supported
by the RGC of Hong Kong (Nos. HKU7044/08P
and HKU7055/09P) and a CRF of Hong Kong.

\end{document}